\documentclass{edp-conf}
\usepackage{graphicx}
\begin{document}
\def\la{\buildrel<\over\sim}
\def\ga{\buildrel>\over\sim}

\TitreGlobal{SF2A 2005}

\title{FAST ROTATION vs. METALLICITY}

\author{Levenhagen, R.S.}\address{Instituto de Astronomia, Geof\'\i sica e 
Ci\^encias Atmosf\'ericas da Universidade de S\~ao Paulo, Brazil}
\author{Leister, N.V.$^1$}
\author{Zorec, J.}\address{Institut d'Astrophysique de Paris, UMR7095 CNRS, 
Universit\'e Pierre \& Marie Curie}
\author{Fr\'emat, Y.}\address{Royal Observatory of Belgium}

\runningtitle{Fast rotation vs. metallicity}

\index{Levenhagen, R.S.}
\index{Leister, N.V.}
\index{Zorec, J.}
\index{Fr\'emat, Y.}

\maketitle

\begin{abstract} Fast rotation seems to be the major factor to trigger the Be 
phenomenon. Surface fast rotation can be favored by initial formation 
conditions such as metal abundance. Models of fast rotating atmospheres and 
evolutionary tracks are used to determine the stellar fundamental parameters 
of 120 Be stars situated in spatially well-separated regions to imply there is
between them some gradient of metallicity. We study the effects of the 
incidence of this gradient on the nature of the studied stars as fast rotators.
\end{abstract}

\section{Introduction}

Ste\c pie\'n (2002) has shown that magnetic fields can spin up early type 
stars in the PMS phase. It acts through mass-accretion and magnetic-disc 
locking, where the efficiency of the interaction can differ according to the 
content of metals in the star and circumstellar environments. Be stars rotate
at $\Omega/\Omega_c\sim0.9$ (Fr\'emat et al. 2005). It is then expected that 
the efficiency of magnetic fields at establishing high initial stellar surface
rotations can be different according to the metallic content of the medium 
where they are formed. 

\section{Method}

 We study whether there is some incidence of the metallicity on setting the Be
phenomenon up by analyzing the age/mass distribution of Be stars situated 
towards the galactic center and in the anti-center direction. The work is 
based on spectroscopic data obtained with FEROS spectrograph at ESO/La Silla 
(Chile) and with the Coud\'{e} spectrograph at the 1.60m telescope of MCT/LNA
(Brazil). The fundamental parameter determination uses models of rotating
stellar interiors and atmospheres according to methods developed in 
Levenhagen (2004), Fr\'emat et al. (2005) and Zorec et al. (2005). 

\section{Results and conclusions}

 Figure~1a shows the distribution of fractional ages ($\tau/\tau_{\rm MS}$ =
age/time spent in the MS) against the stellar mass. Although most of the 
studied stars lay in the second half of the MS strip, a non negligible number
of them is still in the first half of the MS evolutionary phase. A lack of 
stars below $\tau/\tau_{\rm MS} =$ 0.2 is noticeable. Stars with masses $M \ga 
12M_{\odot}$ approach the TAMS limit. This may be due to the fast evolution of 
massive stars and to the lack of massive Be stars at ages $\tau/\tau_{\rm MS} 
\la$ 0.5 because of their rapid loss of angular through high mass-loss rates,
which convert them into low rotators and disable them to display the Be 
phenomenon. Figure~1b shows samples of Be stars divided into ``galactic-center"
and ``anti-center" groups. The ``center" group outnumbers the ``anti-center" 
one, which is at odds with the announced expectancy. However, the distinction 
done here is based on the number of Be stars, while the result would be more 
reliable if we could obtain differences in the fractions of Be/B-type objects 
in the studied space volumes.\par

\begin{figure}[h]
\centering
\includegraphics[height=7cm]{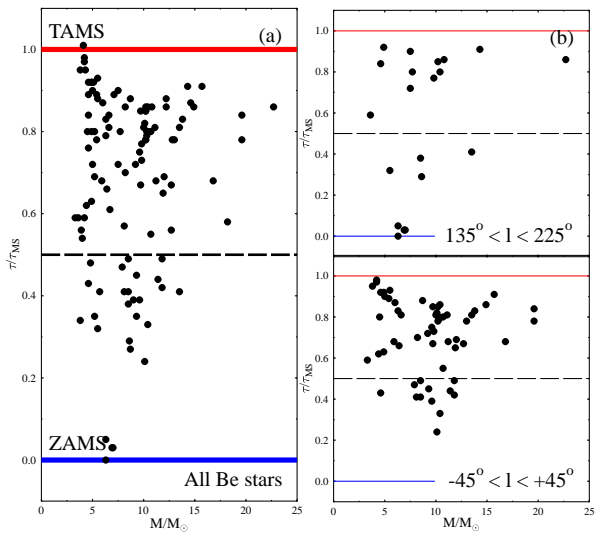}
\caption{(a): Fractional ages $\tau/\tau_{\rm MS}$ against mass of all 
studied Be stars. (b): Same as (a), but for Be stars located towards de 
galactic center and anti-center.}
\label{f1}
\end{figure}

\end{document}